\documentclass[aps,prl,twocolumn,showpacs,showkeys]{revtex4-1}

\usepackage{graphicx}
\usepackage{amsthm,amssymb,amsmath}
\usepackage{textcomp}
\usepackage{xspace}
\usepackage{color}

\newcommand{\ket}[1]{\mbox{$ | #1 \rangle $}}

\newcommand{\tc}{\textdegree \xspace}

\begin{document}

% \preprint{APS/123-QED}

\title{Demonstration of Quantum Nonlocality in presence of Measurement Dependence}

\author{Djeylan Aktas}
\affiliation{Universit\'e Nice Sophia Antipolis, Laboratoire de Physique de la Mati\`ere Condens\'ee, CNRS UMR 7336, Parc Valrose, 06108 Nice Cedex 2, France.
}
\author{Anthony Martin}%
%\email{Second.Author@institution.edu}
\affiliation{Group of Applied Physics, University of Geneva, CH-1211 Geneva 4, Switzerland
}
\author{Gilles P\"utz}%
\affiliation{Group of Applied Physics, University of Geneva, CH-1211 Geneva 4, Switzerland
}

\author{Rob Thew}%
\affiliation{Group of Applied Physics, University of Geneva, CH-1211 Geneva 4, Switzerland
}

\author{S\'ebastien Tanzilli}%
\email{sebastien.tanzilli@unice.fr}
\affiliation{Universit\'e Nice Sophia Antipolis, Laboratoire de Physique de la Mati\`ere Condens\'ee, CNRS UMR 7336, Parc Valrose, 06108 Nice Cedex 2, France.
}

\author{Nicolas Gisin}%
\email{nicolas.gisin@unige.ch}
\affiliation{Group of Applied Physics, University of Geneva, CH-1211 Geneva 4, Switzerland
}

\date{\today}

\begin{abstract}
Quantum nonlocality stands as a resource for Device Independent Quantum Information Processing (DIQIP), as, for instance, Device Independent Quantum Key Distribution.
We investigate experimentally the assumption of limited Measurement Dependence, \textit{i.e.}, that the measurement settings used in Bell inequality tests or DIQIP are partially influenced by the source of entangled particle and/or by an adversary. Using a recently derived Bell-like inequality [\textit{Phys. Rev. Lett. \textbf{113} 190402}] and a $99\%$ fidelity source of partially entangled polarization photonic qubits, we obtain a clear violation of the inequality, excluding a much larger range of measurement dependent local models than would be possible with an adapted Clauser, Horne, Shimony and Holt (CHSH) inequality. It is therefore shown that the Measurement Independence assumption can be widely relaxed while still demonstrating quantum nonlocality.
\end{abstract}

\pacs{03.65.Ud, 03.67.-a, 03.67.Bg, 03.67.Dd, 42.50.Dv, 42.65.Lm}

\keywords{Entanglement, non-locality}

\maketitle

\textit{Introduction--}
The violation of a Bell inequality demonstrates that the observed correlations cannot be explained by any theoretical model based only on local variables that propagate gradually and continuously through space. This seminal result by John S. Bell~\cite{Bell1964} is nowadays considered fundamental for our understanding of quantum physics. Indeed, today the best way to demonstrate that one masters some quantum degree of freedom of some physical system is to violate a Bell inequality using this degree of freedom. Additionally, quantum nonlocality, \textit{i.e.}, the violation of a Bell inequality, is the resource physicists and computers scientists exploit in Device Independent Quantum Information Processing (DIQIP), like Device Independent Quantum Key Distribution~\cite{Pironio_2009,Vazirani2014} and Device Independent Quantum Random Number Generators~\cite{Pironio2010}. This dual fundamental-\&-applied importance of quantum nonlocality has triggered an interesting scientific race to close both the locality and the detection loopholes simultaneously in one single experiment~\cite{Hofmann2012,Giustina2013,Volz2006,Christensen2013,NIST}.

\begin{figure}
\includegraphics[width=\columnwidth]{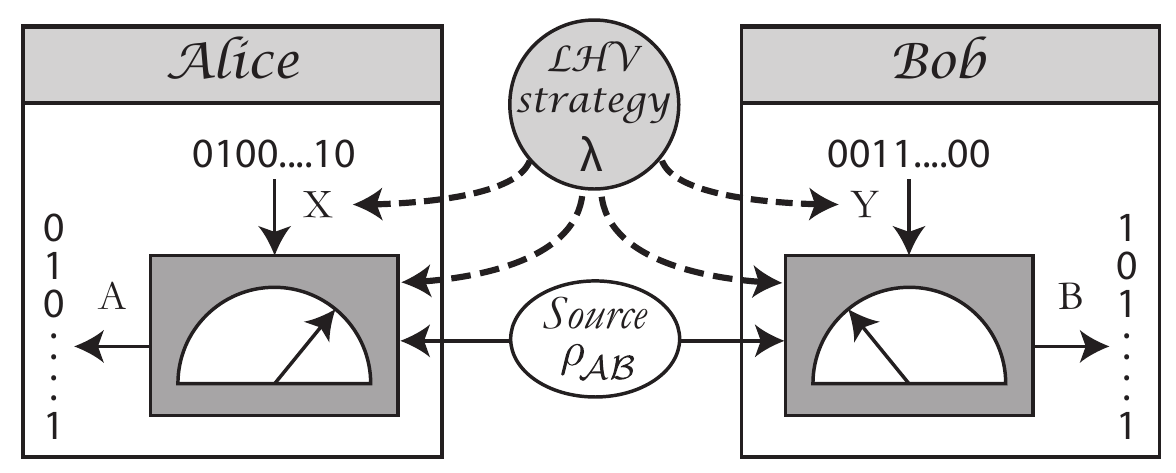}
\caption{\label{fig_principle} Schematic of a standard quantum correlation measurement scheme, established in the presence of a local hidden variable (LHV) strategy. The two users, Alice ($\mathcal{A}$) and Bob ($\mathcal{B}$), each have a measurement apparatus. These devices each take a binary input (x,y) and return a binary output (a,b). They can also be provided with a hidden common variable, $\lambda$, to mimic a non-local quantum resource. Note that the LHV can influence the input choices of both Alice and Bob. This scenario is called measurement dependent locality.}
\end{figure}

However, there is another sort of assumption in the derivation of Bell inequalities, that of Measurement Independence (also known as ``Freedom of Choice'', ``Free Randomness'', or sometimes loosely denoted by ``Free Will''). This assumption states that the hypothetical local variable, traditionally referred to as $\lambda$, does not influence the local choices of measurement settings performed by the two parties Alice and Bob. Formally, denoting by $x$ and $y$ the measurement settings of Alice and Bob, respectively, this assumption reads~\cite{Hall2011,Barrett2011}:
\begin{align}
P(xy|\lambda)=P(xy) \text{ }\forall x,y,\lambda.
\end{align}
This is a very natural assumption. Indeed, if the local variable $\lambda$ and the measurement settings $x,y$ would be fully correlated, then there would be some sort of cosmological conspiracy, sometimes called hyper-determinism, where everything, somehow, was set-up at the big-bang; an admittedly not very interesting assumption. But what about intermediate cases where $\lambda$ partially influences the outcomes of the random number generators that in realistic Bell experiments determine the measurement settings $x$ and $y$ (see Fig. \ref{fig_principle})? This is especially relevant in the context of DIQIP where the influence could be due to an active adversary.

Several ways of relaxing the assumption of Measurement Independence have already been pursued~\cite{Hall2011,Barrett2011,Colbeck2011,Koh2012,Thinh2013,Grudka2013}. Recently some of us, together with collaborators, followed a different path, where we assumed a limited Measurement Dependence of the form~\cite{Puetz_PRLBell_2014}:
\begin{align}
\label{LMD}
0<\ell\leq P(xy|\lambda).
\end{align}
This means that even conditioned on the local variable $\lambda$, every input pair $(x,y)$ can occur in each run of the experiment with at least a probability $\ell$.

With this mild assumption we could prove that there exist quantum correlations that are nonlocal for all $\ell >0$~\cite{Puetz_PRLBell_2014}. Moreover, this result was obtained using a generalized Bell-like inequality well suited for experimental tests. It is the purpose of this letter to demonstrate an experimental violation of this inequality. Note that if one would stick to the CHSH-Bell inequality~\cite{Clauser1969}, the smallest $\ell$ one could tolerate theoretically would be only~\cite{Puetz_PRLBell_2014} $\ell\geq\frac{2-\sqrt{2}}{4}\approx 0.146$. Our experiment, however, lowers this bound on Measurement Dependence down to 0.090. Consequently, our work excludes a much larger range of measurement dependent local models than would be possible using the CHSH inequality (while still using only binary inputs and outcomes).

Formally, a correlation $P(abxy)$ is \textit{$\ell$-measurement dependent local} (MDL) iff it can be written as
\begin{align}
P(abxy)=\int d\lambda P(\lambda)P(xy|\lambda)P(a|x\lambda)P(b|y\lambda),
\end{align}
where $P(xy|\lambda)$ satisfies Eq.~(\ref{LMD}) and $P(abxy)$ denotes the joint probability distribution of results $a$ \& $b$ and measurement settings $x$ \& $y$ on Alice's and Bob's sides, respectively.

It has been shown in Ref.~\cite{Puetz_PRLBell_2014} that all $\ell$-measurement dependent local correlations fulfill the Bell-like inequality:
\begin{align}
\ell P(0000)- (1-3\ell)\big(P(0101) + P(1010)& + P(0011)\big)\nonumber\\
\label{Eq_ineqfull}
& \stackrel{MDL}{\leq} 0.
\end{align}
In the same way that a violation of the CHSH inequality excludes all possible local models, a violation of inequality (\ref{Eq_ineqfull}) for a given $\ell$ excludes all possible $\ell$-measurement dependent local models. If we make the additional assumption that, for an observer that does not have access to the local hidden variable $\lambda$, the measurement settings seem distributed fairly, i.e. $P(xy)=\sum_\lambda P(\lambda)P(xy|\lambda)=\frac{1}{4}$, then inequality (\ref{Eq_ineqfull}) becomes
\begin{align}
\ell P(00|00)- (1-3\ell)\big(P(01|01) + P(10|10)& + P(00|11)\big)\nonumber\\
\label{Eq_ineq}
& \stackrel{MDL}{\leq} 0,
\end{align}
where $P(ab|xy)$ is the conditional probability of getting outcomes $a$ for Alice and $b$ for Bob if the inputs were $x$ and $y$, respectively.

\textit{Quantum state engineering and experimental realization--}
To highlight the measurement dependent non-locality of quantum physics for any $\ell>0$, we need, first to prepare a pure 2-qubit non-maximally entangled state~\cite{White_Explore_2001} of the form~\cite{Puetz_PRLBell_2014}:
\begin{equation}
\label{Eq_State}
|\Psi\rangle = \frac{1}{\sqrt{3}}\left( \frac{\sqrt{5}+1}{2} |0_\mathcal A,0_\mathcal B \rangle +  \frac{\sqrt{5}-1}{2} |1_\mathcal A,1_\mathcal B\rangle\right).
\end{equation}
This state with the Golden ratio as Schmidt coefficient is the quantum state that leads to the largest violation of inequality (\ref{Eq_ineq}).
Next, on Alice's ($\mathcal A$) and Bob's ($\mathcal B$) sides, respectively, we need to apply the following projective measurements: 
\begin{equation}\label{Eq_settings}
\begin{array}{l}
\mathcal A
\left\{
\begin{array}{l}
\ket {A_0(\theta)} = \cos(\theta) \ket 0 + \sin(\theta) \ket 1\\
\ket {A_1(\theta)} = \ket {A_0(\theta+\pi/4)} 
\end{array}
\right.\\
\mathcal B
\left\{
\begin{array}{l}
\ket {B_0(\theta)} = \ket {A_0(-\theta)}\\
\ket {B_1(\theta)} = \ket {A_1(-\theta)},
\end{array}
\right.
\end{array}
\end{equation}
with $\theta = \rm acos \sqrt{1/2-1/\sqrt{5}}\approx 76.71$ degrees.

\begin{figure}[h!]
\resizebox{1\columnwidth}{!}{\includegraphics{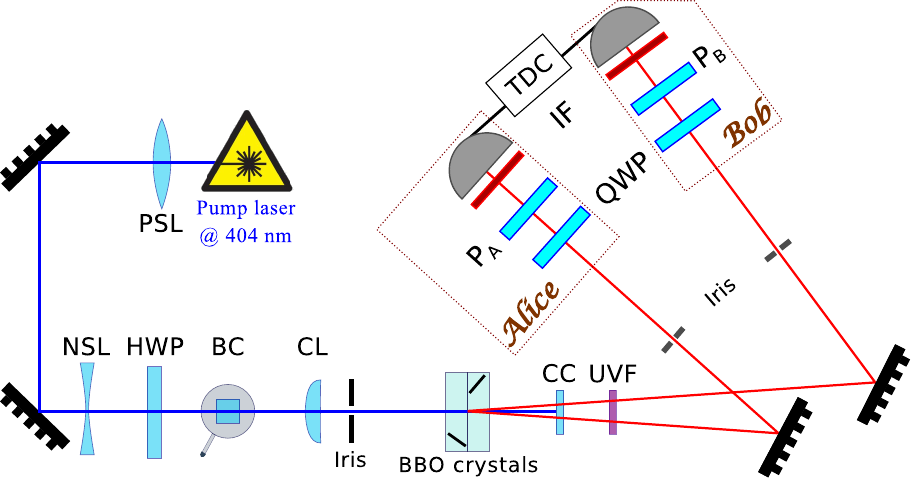}}
\caption{Schematic of the source. A laser at 404\,nm pumps two cascaded BBO crystals generating pairs of polarization entangled photons by SPDC at 808\,nm~\cite{Kwiat_Source2crystalsPRL_1995}. The half-wave plate (HWP), placed after a set of positive and negative spherical lenses (PSL, NSL), associated with a birefringent Crystal (BC) serves to adjust both the polarization state and the shape of the pump laser (in front of the BBO crystals). By setting its angle appropriately, it is possible to produce arbitrary pure two-photon states, \textit{i.e.}, to choose any desired values of weights of the coherent superposition $\ket{VV}$ and $\ket{HH}$, as well as the relative phase between these two contributions. The photon pairs generated by the paired BBO crystals then pass through a compensation crystal (CC) to erase any distinguishability between the two non-linear processes involved. Subsequently, the combination of a pinhole and a single mode fiber ensures proper spatial filtering in each arm. Moreover, a set of ultraviolet notch (UVF) and infrared band pass filters (IF) are employed to reject the pump photons. The amount of entanglement is then analyzed via correlation measurements. This is achieved by employing standard polarization state analyzers composed, at each location, of a quarter-wave plate (QWP), a polarizer $P_{A,B}$, and one single photon detector (silicon avalanche photodiodes (Si-APD), Excelitas SPCM-AQR-16-FC). The signals out of the APDs are sent to a time-to-digital converter (TDC, IDQ-800) to record the coincidence counts. CL: cylindrical lens.}
\label{Fig_source}
\end{figure}

To produce this state coded on the polarization modes of two photons, we employ the source depicted in \figurename{~\ref{Fig_source}}~\citep{Kwiat_Source2crystalsPRL_1995,Steinlechner2012} and identify the qubit state $\ket{0}$ with V-polarization and $\ket{1}$ with H-polarization. Starting with a pump laser at 404\,nm, pairs of polarization entangled photons are produced at 808\,nm by spontaneous parametric down-conversion (SPDC) in a cascade of two type-I beta-barium-borate (BBO) crystals having orthogonal optical axes.
The type-I phase matching in the first (resp. second) crystal produces the state $\ket{V_\mathcal A, V_\mathcal B}$ (resp.$\ket{H_\mathcal A, H_\mathcal B} $) when pumped by a vertically (resp. horizontally) polarized laser beam. In this way, rotating the linear polarization of the pump beam in front of such a two-crystal configuration can generate any desired state of the form $c_V \ket{V_\mathcal A, V_\mathcal B}+c_H \ket{H_\mathcal A, H_\mathcal B}$. This is made possible provided the two cascaded non-linear processes, associated with both filtering and compensation stages, produce perfectly indistinguishable photon pairs in all other degrees of freedom. Here, $c_V$ and $c_H$ denote the probability amplitudes associated with the two contributions to the state. The strength of the source lies in the flexibility to produce states ranging from product states (pump beam fixed either with an horizontal or vertical polarization state) to maximally entangled states (pump beam polarized at 45\textdegree). Consequently, generating the state of Eq.~(\ref{Eq_State}) amounts to choosing the polarization state for the pump laser oriented at 20.9\textdegree. This is achieved with a half-wave plate (HWP) placed in between the laser and the two non-linear crystals (see \figurename{~\ref{Fig_source}} and related caption for more details).

\begin{figure}
\begin{tabular}{c c}
Ideal state & Generated state\\
\resizebox{0.5\columnwidth}{!}{\includegraphics{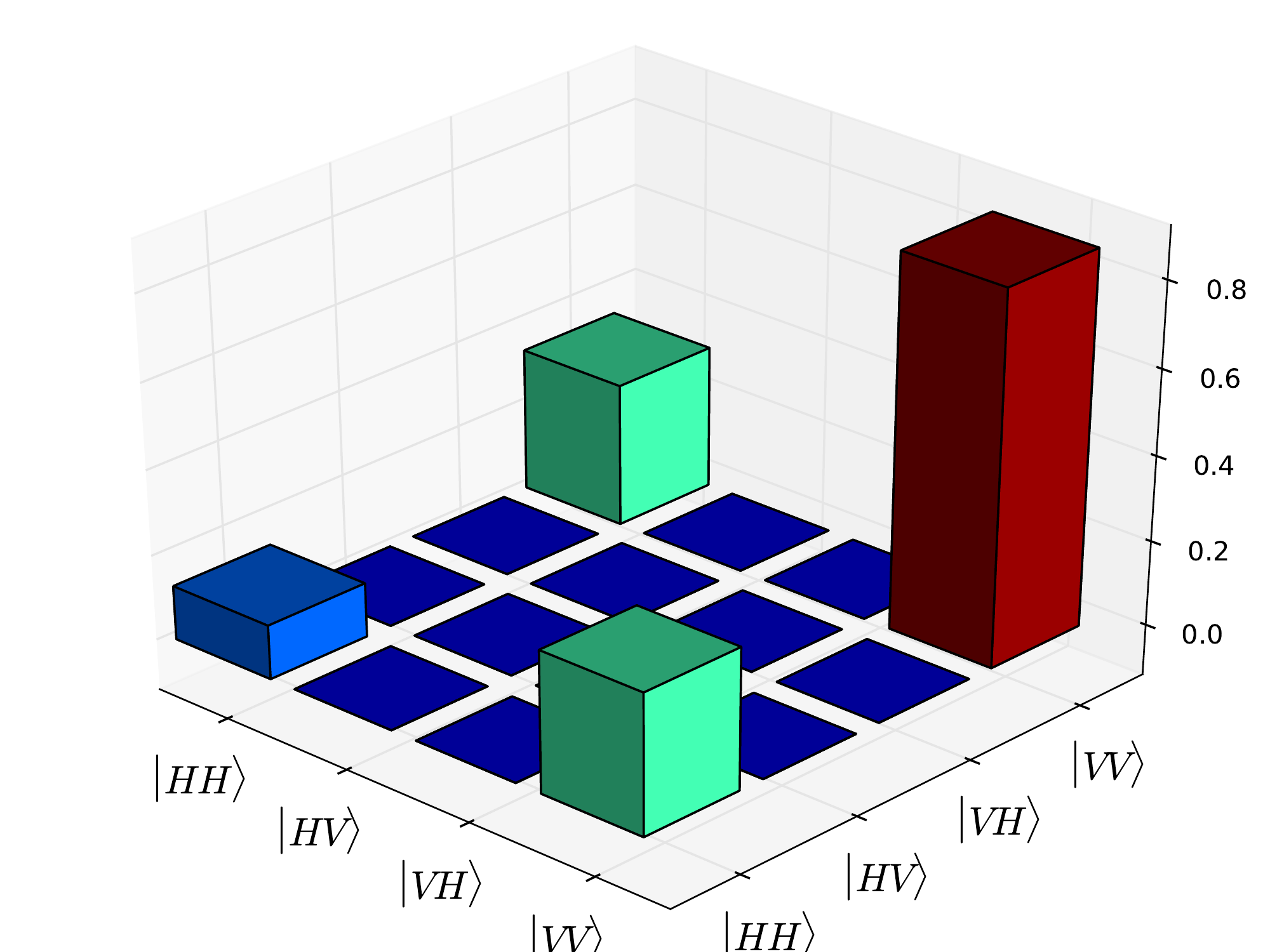}} & \resizebox{0.5\columnwidth}{!}{\includegraphics{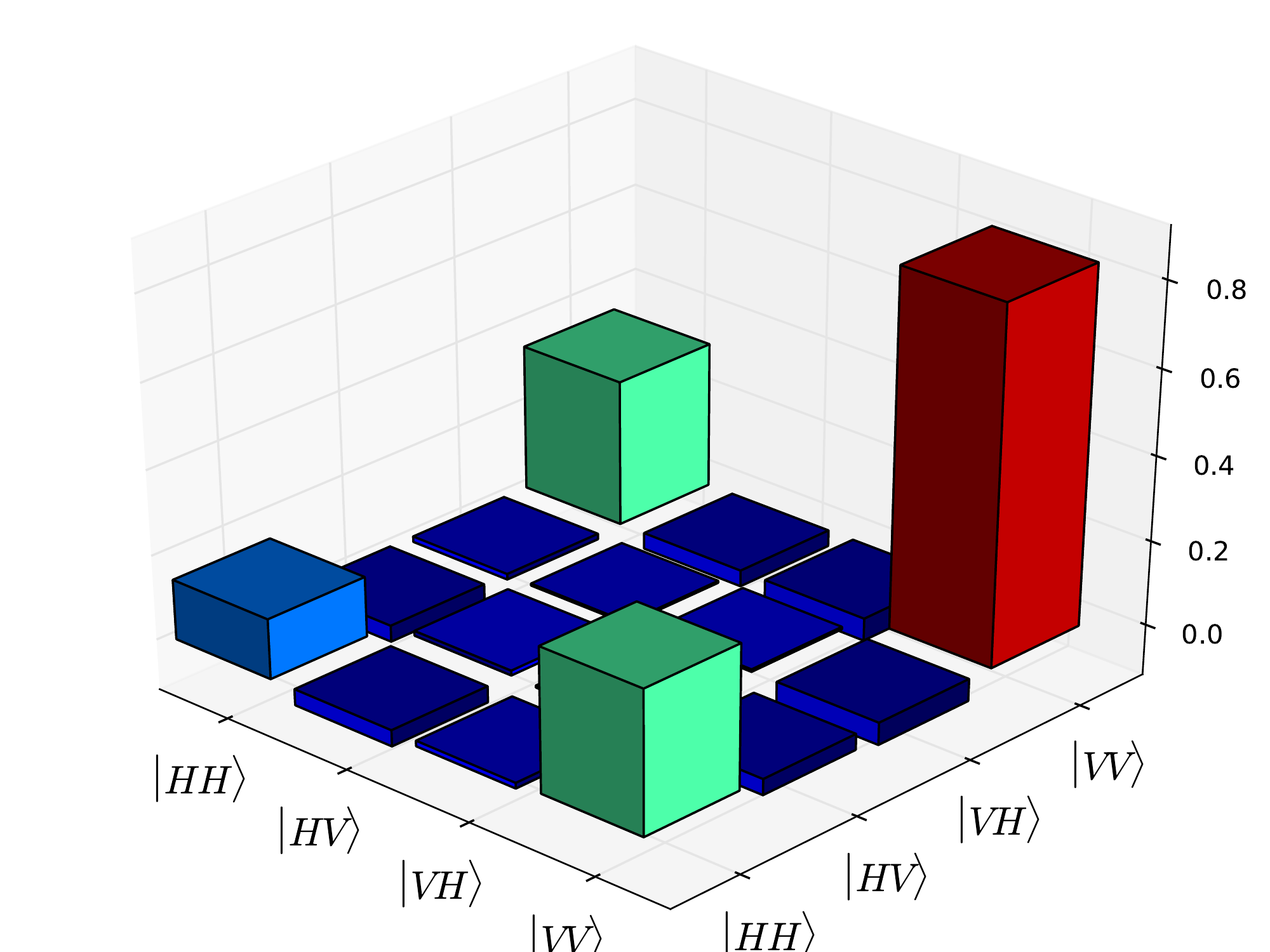}}\\%
(a) & (c)\\
\resizebox{0.5\columnwidth}{!}{\includegraphics{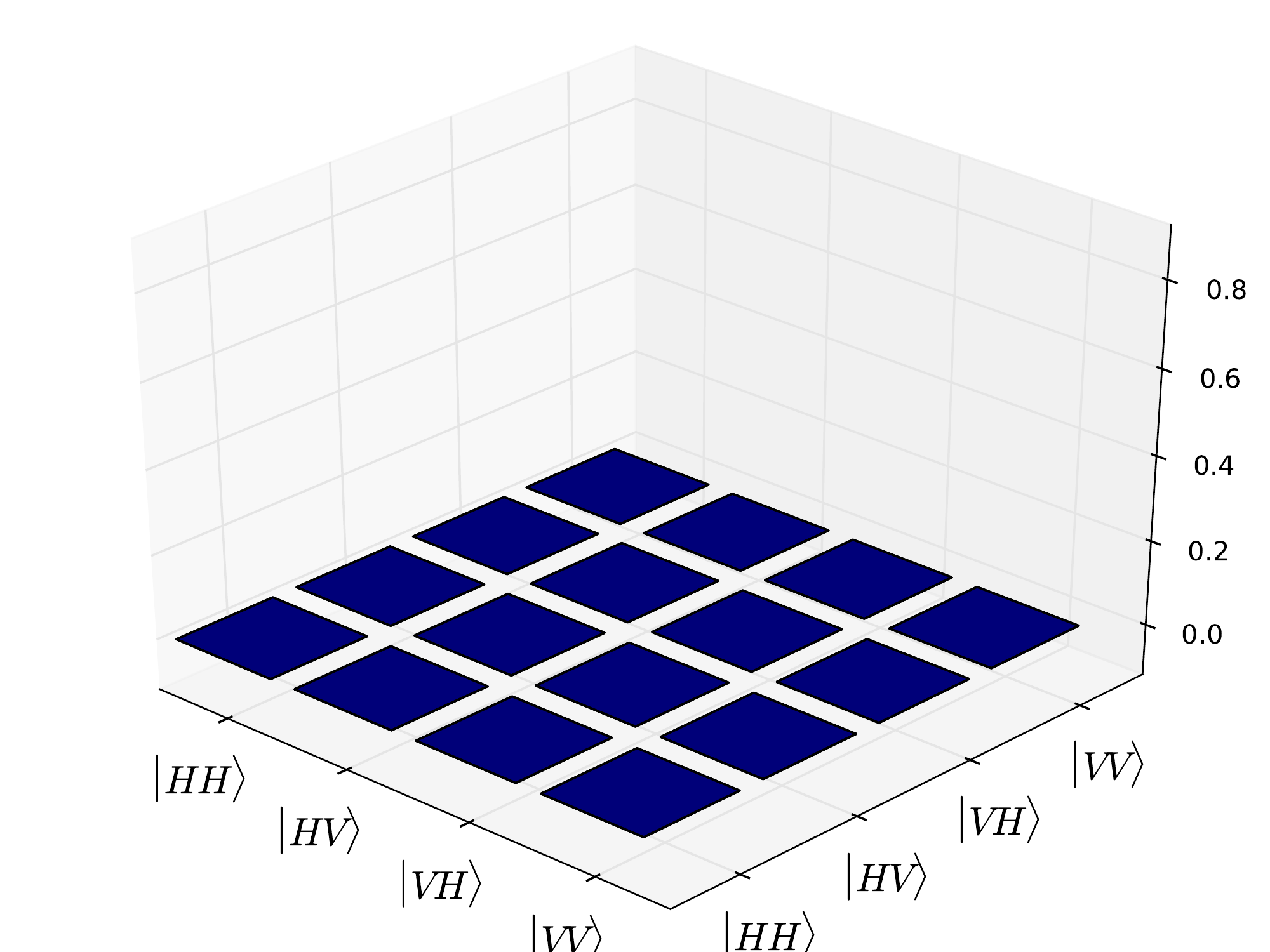}} & \resizebox{0.5\columnwidth}{!}{\includegraphics{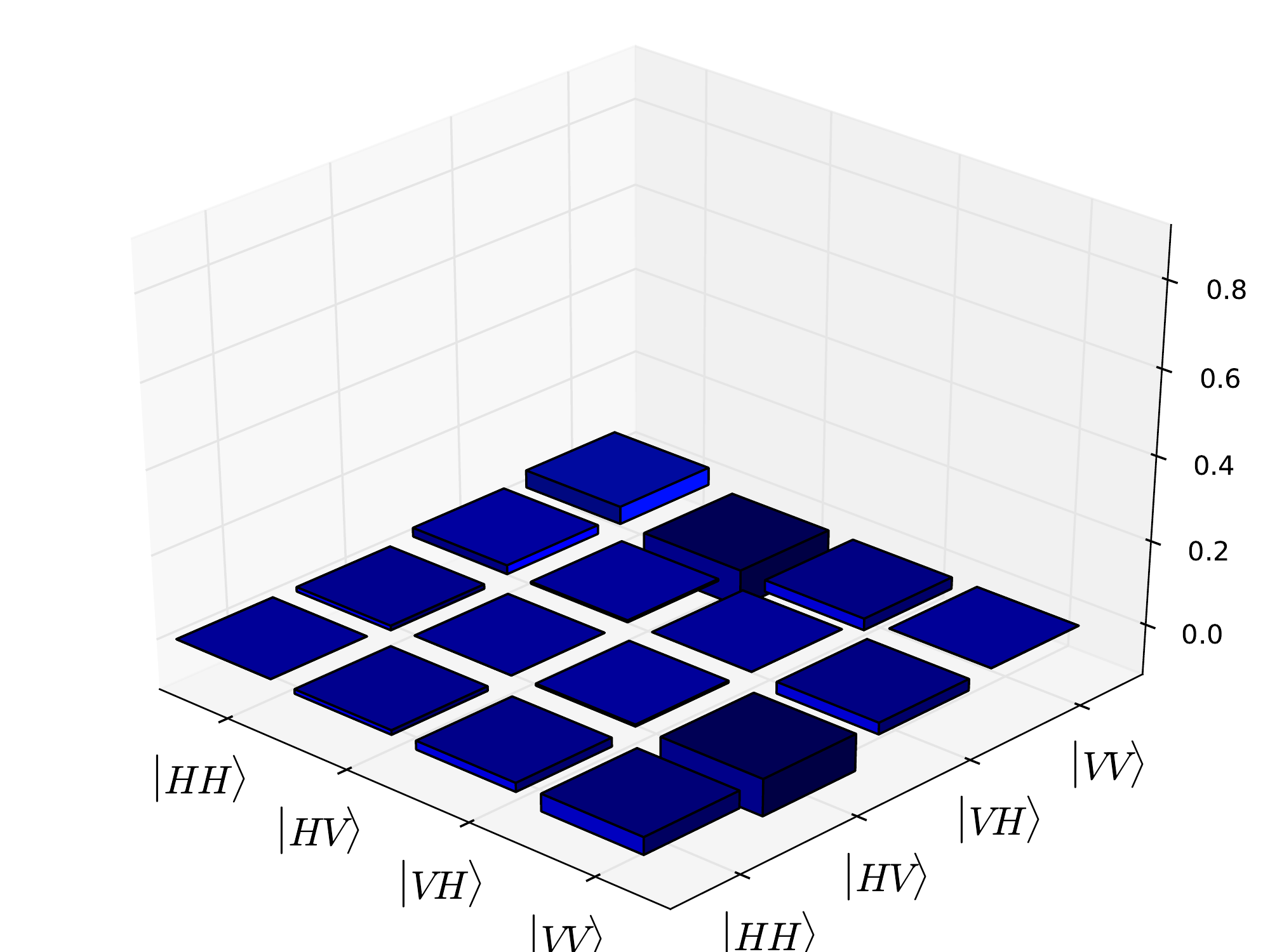}}\\
(b) & (d)
\end{tabular}
\caption{\label{Fig_states}Real and imaginary parts of the density matrices associated with the ideal (a \& b) and the produced (c \& d) states.
We obtain a fidelity of 0.99(1) with the target state.}
\end{figure}

Once the experimental parameters are suitably fixed, we first perform a full quantum tomography of the produced state from which we infer the fidelity to the ideal state of Eq.~(\ref{Eq_State}). To this end, we follow the procedure outlined in Ref.~\cite{James_Qubit_2001}. The corresponding results are given in \figurename{~\ref{Fig_states}}, where (a) and (b) represent the real and imaginary parts of the density matrix associated with the state of Eq.~(\ref{Eq_State}), respectively, whereas (c) and (d) are those of the produced and subsequently characterized state. The corresponding fidelity between the target state and the produced state is measured to be 0.99(1).

\textit{Results--} The obtained results on the correlation measurements are summarized in Table~\ref{Table_results_summary}.
The reported data are recorded using a standard coincidence technique based on two single photon detectors connected to the start and stop inputs of a time-to-digital converter (TDC, see \figurename{~\ref{Fig_source}}). The corresponding coincidence (third column) and noise (fourth column) values are registered over 30\,s integration time, inside and outside the obtained coincidence peak (not represented), respectively. Note that by noise we refer to accidental coincidence events that are mainly due to the detectors' dark counts. Then, the probabilities (last two columns) are inferred by normalizing, for a given measurement, the corresponding coincidence value by the total number of coincidences obtained for all the possible settings in the considered basis. Finally, the 'net' probabilities correspond to similar normalization with noise figure discarded.

\begin{table*}[t]
   \centering
\begin{tabular}{|c|c|c|c|c|c|c|}
         \hline  
         Alice's setting $x$ & Bob's setting $y$ & Raw Coincidences (/30\,s) & Noise (/30\,s)  & $P^{\rm raw}(ab|xy)$ & $P^{\rm net}(ab|xy)$ \\\hline  
         $0$ & $0$ & 2939 / 35183 & 14 / 269 &{$P(00|00)=0,0835(10)$} & {$P(00|00)=0,0838(15)$} \\
         $0$ & $1$ &129 / 36658 & 26 / 270 &$P(01|01)=0,0035(3) $& $P(01|01)=0,0028(3) $\\
         $1$ & $0$ & 114 / 34693 & 32 / 280 &$P(10|10)=0,0033(3)$ & $P(10|10)=0,0024(3)$\\
         $1$ & $1$ & 130 / 36962 & 23 / 276 &$P(00|11)=0,0035(3) $& $P(00|11)=0,0027(3)$\\
         \hline
\end{tabular}
\caption{\label{Table_results_summary}Experimental coincidence values using all the necessary settings as outlined in Ref.~\cite{Puetz_PRLBell_2014}. Four measurement settings ($\alpha, \beta$), ($\alpha, \beta_{\perp}$), ($\alpha_{\perp}, \beta$), ($\alpha_{\perp}, \beta_{\perp}$) give the probabilities $P(00|00)$, $P(01|01)$, $P(10|10)$ and $P(11|00)$ for the analysis of inequality (\ref{Eq_ineq}). Here, $a$ and $b$ refer to the outputs on Alice's and Bob's sides respectively, while $x$ and $y$ label the inputs. For all the bases, in the column labeled 'Coincidences', the left values correspond to the number of raw coincidence events recorded over a 30\,s integration time for the given setting, while the right values corresponds to the total number of raw coincidence events recorded over a 30\,s integration time for all the possible settings in the considered basis.
Moreover, the column 'Noise' gives the number of accidental events recorded over 30\,s. The associated probabilities $P^{\rm raw}(ab|xy)$ and $P^{\rm net}(ab|xy)$ are computed by normalizing a given coincidence value (left side of the third column) by the total number of acquired coincidences in the same basis (right side of the third column), without (raw) and with (net) noise discarded, respectively.}
\end{table*}

Those coincidence and noise values are given as a function of Alice and Bob's respective settings outlined in Eq.~(\ref{Eq_settings}).
To perform this projection in our setup, the two QWPs (see \figurename{~\ref{Fig_source}}) are set to 0\tc and the polarizers are set to $\alpha$=13.3\textdegree\,and $\alpha_{\perp}$=103.3\tc for $A_0$, $\alpha$=58.3\tc and $\alpha_{\perp}$=328.3\tc for $A_1$, $\beta$=-13.3\tc and $\beta_{\perp}$=76.7\tc for $B_0$, $\beta$=31.3\tc and $\beta_{\perp}$=301.3\tc for $B_1$.
The recorded coincidence values in Table~\ref{Table_results_summary}, together with the total number of coincidences, are used to compute the four probabilities $P(ab|xy)$ necessary to evaluate the terms of the MDL inequality (\ref{Eq_ineq}). As mentioned above, a violation of this inequality for a given $\ell$ implies that no measurement dependent local model with $P(xy|\lambda)\geq\ell$ can reproduce the correlation.

Let us emphasize that in our experiment both the detection and the locality loopholes remain widely open, as here we concentrate on Measurement Dependence. In particular we did not change the measurement settings between each run of the experiment, but only in-between successive series of runs, as is often done in Bell tests.

Accordingly, we can exclude all values of $\ell$ for which our data violates the MDL inequality, with $\ell_{\rm net}>0.074(6)$ or $\ell_{\rm raw}>0.090(4)$, where the indices $\rm net$ and $\rm raw$ refer to the net (noise discarded) and raw (noise included) recorded data, respectively. In theory one can demonstrate quantum nonlocality for any $\ell>0$. The main experimental limitation comes from several small imperfections, notably the non-zero multiple photon-pair generation, non-ideal measurement apparatus, noise coming from the detectors and external photons, as well as non-unit generated state fidelity. Let us recall that a maximal violation of a CHSH MDL-adjusted inequality only excludes $\ell>0.1465$~\cite{Puetz_PRLBell_2014,Clauser1969}. 

\textit{Conclusion--} One important assumption in Bell inequalities, including the well known CHSH-Bell inequality, used to prove the non-locality aspect of a quantum state, is that of Measurement Independence, that is the choices of the measurement settings of Alice and/or Bob are assumed to be not influenced by any kind of external and classical entity related to the source of entangled quantum particles. Typically, such an influence could be due to an active adversary that twisted the random number generator used in applications of Device-Independent Quantum Information Processing (DIQIP). If we suppose that the adversary has full control, i.e. $\ell=0$, then it becomes impossible to prove the non-local aspect of the measured correlations and thus all DIQIP applications become impossible. The question that arises concerns the assumption of having only partial measurement dependence. To address the possibility of such influences, the experiment presented in this letter addresses this natural question by demonstrating the violation of the inequality introduced in Ref.~\cite{Puetz_PRLBell_2014}. By exploiting a specific non-maximally entangled two-photon state associated with suitable measurement settings, we have demonstrated that the produced state is non-local with a much less restrictive assumption on measurement dependence~\footnote{This can be translated to limited minimal detection efficiency $\eta_{min}$ [in preparation] in which case our data lead to $\eta_{min}^{net}=\sqrt{\ell_{net}}\approx 0.27$ and  $\eta_{min}^{raw}=\sqrt{\ell_{raw}}\approx 0.30$} than would be necessary with standard Bell-CHSH approaches. 

\textit{Acknowledgments}-- The authors would like to acknowledge L. Labont\'e for his help in the data acquisition process, B. Sanguinetti for fruitful discussions on the laser system, T. Barnea for discussions on the theoretical part, the Fondation iXCore pour la Recherche, the Fondation Simone \& Cino Del Duca (Institut de France), the European Chist-Era project DIQIP and the Swiss Project NCCR:QSIT for financial support, as well as QuTools for technical support.

%\bibliography{Aktas_PRL_MArch2015}

\end{document}